\documentclass[preprintnumbers,amsmath,amssymb,floatfix,10pt,prd,onecolumn,
superscriptaddress,nofootinbib]{revtex4}
\usepackage{latexsym,float}
\usepackage{epsfig,color}
\usepackage{epstopdf}
\usepackage{amssymb}
\usepackage{verbatim}
\usepackage{multirow}

\begin{document}

\title{\bf Dynamical Analysis of Cylindrically Symmetric Anisotropic Sources in $f(R,T)$
Gravity}

\author{M. Zubair}
\email{mzubairkk@gmail.com;
drmzubair@ciitlahore.edu.pk}\affiliation{Department of Mathematics,
COMSATS Institute of Information Technology Lahore, Pakistan}

\author{Hina Azmat}
\email{hinaazmat0959@gmail.com}\affiliation{Department of
Mathematics, COMSATS Institute of Information Technology Lahore,
Pakistan}

\author{Ifra Noureen}
\email{ifra.noureen@gmail.com;
ifra.noureen@umt.edu.pk}\affiliation{Department of Mathematics,
University of Management and Technology, Lahore, Pakistan.}

\date{\today}

\begin{abstract}
In this paper, we have analyzed the stability of cylindrically
symmetric collapsing object filled with locally anisotropic fluid in
$f(R,T)$ theory, where $R$ is the scalar curvature and $T$ is the
trace of stress-energy tensor of matter. Modified field equations
and dynamical equations are constructed in $f(R,T)$ gravity.
Evolution or collapse equation is derived from dynamical equations
by performing linear perturbation on them. Instability range is
explored in both Newtonian and post-Newtonian regimes with the help
of adiabetic index, which defines the impact of physical parameters
on the instability range. Some conditions are imposed on physical
quantities to secure the stability of the gravitating
sources.
\end{abstract}

\maketitle

\section{Introduction}

Astrophysics and theories regarding gravity bear two emerging issues:
aftermath of gravitational collapse and explorations regarding stability of
celestial bodies. Collapse of a star depends on availability of its fuel,
exhaustion of all of its fuel makes the gravitational collapse indispensable
because inward gravitational pull, meanwhile overpowers the outward drawn
force \cite{1}. Undoubtedly, the size of collapsing star determines the end
state of evolution, the life span of huge stars having mass equivalent to ten
to twenty solar masses is not comparable to that of stars assuming
sufficiently little mass. Moreover, the more massive stars are the more
vulnerable to instability due to heat flux emanating because of high energy
dissipation during collapsing phenomenon \cite{2,3}.

The astronomical bodies seek attention only if they show resistance
against fluctuations and remain stable. In 1964, Chandrasekhar
\cite{4} presented investigations of primary level on dynamical
stability of spherical bodies. He pinpointed the instability range
of a star assuming mass `$M$' and radius `$r$' with the help of
adiabatic index $\Gamma$ comprising the inequality $\Gamma
\geq{\frac{4}{3}+n\frac{M}{r}}$. Adiabatic index is a tool which is
defined as pressure to density ratio and used to determine the
stability range of celestial objects. It imposes some conditions on
physical parameters to meet stability criterion which is defined for
stellar objects. Instability problems in the theory of general
relativity (GR) coupled with dissipation, shear, zero expansion,
radiation, isotropy and local anisotropy was addressed by Herrera
and his companions in \cite{5}-\cite{10}. They established the fact
that instability range is vulnerable to drastic changes if slight
variations take place in isotropic profile and shearing effects.
Sharif and Abbas \cite{11} analyzed the collapse of charged
cylindrical celestial bodies for non-adiabatic and perfect fluid.
Sharif and Azam \cite{12} presented stability analysis for
cylindrically symmetric thin-shell wormholes.

Researchers contributed substantially in addressing the instability
problems with the help of GR but it is no longer helpful to explain
the cosmological and astrophysical phenomena in a satisfactory way
in the presence of dark matter. The concept of exotic matter like
dark energy and observational evidences of expanding universe has
put the theoretical cosmology into crisis. Due to limitations of GR
on large scales, gravitational modified theories grab the attention
of astrophysicists. With the help of these theories, they put great
efforts to analyze collapsing phenomenon and stability of
astronomical bodies. Among different modified theories, $f(R)$
gravity admits one of the basic modifications in Einstein-Hilbert
action which includes high order curvature $R$ to explain the above
mentioned exotic matter. With the help of various observations like
cosmic microwave background, clustering spectrum and weak lensing
\cite{13}-\cite{16}, it was concluded that the stability range is
increased by the inclusion of curvature terms of higher order.

In modified theories of gravity, collapsing phenomenon has been
widely studied. Collapse of self gravitating dust particles has been
discussed in \cite{17}, where authors found that analysis of
gravitational collapse is an important tool to constrain the
modified models that present late time cosmological acceleration.
Meanwhile, Gosh and Maharaj \cite{18} established exact solutions of
null dust non-static cluster of particles in $f(R)$ gravity,
constrained by constant curvature describing anti de-sitter
background. Some highly important prospects of celestial collapse
for $f(R)$ theory are worked out in \cite{19}-\cite{22}. Sharif and
Bhatti \cite{23} discussed instability conditions for cylindrically
symmetric self-gravitating objects surrounding in charged
expansion-free anisotropic environment. Kausar and Noureen \cite{24}
discussed the evolution of gravitating sources in the context of
$f(R)$ theory and concluded that adiabatic index has its dependence
upon the electromagnetic background, mass and radius of spherically
symmetric bodies.

In 2011, another modification to the theory of GR was introduced by
Harko et al. in \cite{25}, which is extension of $f(R)$ theory, such
theory is named as $f(R,T)$ theory. The $f(R,T)$ theory of gravity
covers curvature and matter coupling and its action includes an
arbitrary function of Ricci scalar `$R$' and trace of energy
momentum tensor $T$. After its introduction, this gained significant
attention and authors discussed its various properties including
reconstructions schemes, energy conditions, cosmological and
thermodynamical implications, neutron stars, scalar perturbations,
wormholes and analysis of anisotropic universe models and stability
etc. in \cite{26}. Shabani and Farhoudi \cite{27} applied dynamical
system approach to elaborate weak field limit and presented analysis
of cosmological implications of $f(R,T)$ models with various
cosmological parameters like Hubble parameter, equation of state
parameter, snap parameter.

Recently, dynamical analysis of self gravitating sources has been
discussed in $f(R,T)$ theory. Noureen and Zubair \cite{28} discussed
dynamical instability of spherically symmetric collapsing star in
the presence anisotropic fluid. The implications of shear free and
expansion free conditions on dynamical instability are also
discussed in the framework of $f(R,T)$ \cite{29}. Motivating from the
significance of non-spherical symmetries, impact of
axially symmetric gravitating sources has also been explored in context of $f(R)$ and $f(R,
T)$ \cite{30}. Yousaf and Bhatti \cite{31} identified dynamical
instability conditions of a self gravitating cylindrical object in
$f(R,T)$ theory of gravity.

In this paper, we have chosen the model ``$f(R,T)=R+\alpha
R^2+\gamma R^n+\lambda T$'' to present the dynamical analysis of
cylindrically symmetric object. In order to present this analysis we
employ perturbation approach on collapse equations and explore the
instability range of the model under consideration with the help of
adiabatic index $\Gamma$ in both Newtonian and post-Newtonian
regimes. The paper has been organized as follows: The next section
contains modified field equations and dynamical equations. In
section \textbf{III}, we have provided the adopted $f(R,T)$ model
and presented perturbation scheme along with corresponding collapse
equation. Discussion about the instability in the form of adiabatic
index in both Newtonian and post-Newtonian regimes is also presented
in the same section. The last section \textbf{IV} concludes our main
findings which is followed by Appendix \textbf{A}.

\section{Dynamical Equations in $f(R,T)$}

The $f(R,T)$ modification of Einstein-Hilbert action is given by

\begin{equation}\label{1}
\int dx^4\sqrt{-g}[\frac{f(R, T)}{16\pi G}+\mathcal{L} _ {(m)}].
\end{equation}
Here, action due to matter is described by $\mathcal{L} _ {(m)}$,
whose different choices can be taken into account, each of which leads to a particular
form of fluid.

The variation of above modified action with respect to metric $g_{\alpha\beta}$
leads to the following set of field equations
\begin{eqnarray}\nonumber
&&R_{\alpha\beta} f_R(R,T)-\frac{1}{2}g_{\alpha\beta} f(R,T)
+(g_{\alpha\beta}\Box-\nabla_{\alpha}\nabla_{\beta})f_R(R,T)
\\\label{2}&&=8\pi GT_{\alpha\beta}^{(m)}-f_T(R,T)T_{\alpha\beta}^{(m)}
-f_T(R,T)\Theta_{\alpha\beta},
\end{eqnarray}
where $f_R(R,T)=\frac{\partial f(R,T)}{\partial R}$, $f_T(R,T)=\frac{\partial f(R,T)}{\partial T}$,
while $\nabla_{\beta}$ and $\Box$ are
derivative operators and represent covariant derivative
and four-dimensional Levi-Civita covariant derivative respectively.
The term $\Theta_{\alpha\beta}$ is defined as
\begin{eqnarray}\label{3}
\Theta_{\alpha\beta}=\frac{g^{\mu\nu}\delta T_{\mu\nu}}{\delta
g^{\alpha\beta}}=
 -2T_{\alpha\beta}+g_{\alpha\beta}\mathcal{L} _ {(m)}-2g^{\mu\nu}\frac{\partial^2\mathcal{L} _ {(m)}}
 {\partial g^{\alpha\beta}\partial g^{\mu\nu}}.
\end{eqnarray}
Here, we have chosen $\mathcal{L} _ {(m)}= \mu$, $8\pi G = 1$, then the expression $\Theta_{\alpha\beta}$ becomes
\begin{eqnarray}\label{4}
\Theta_{\alpha\beta}=-2T_{\alpha\beta}+\mu g_{\alpha\beta}.
\end{eqnarray}
For aforementioned choice of matter Lagrangian and Eq.(\ref{3}), the
modified field equations (\ref{2}) will be as follows:
\begin{eqnarray}\label{5}
G_{\alpha\beta}&=& T_{\alpha\beta}^{eff},
\end{eqnarray}
where
\begin{eqnarray}\label{6}
T^{eff}_{\alpha\beta}&=&\frac{1}{f_R}\left[(f_T+1)T^{(m)}_{\alpha\beta}-\mu
g_{\alpha\beta}f_T+
\frac{f-Rf_R}{2}g_{\alpha\beta}(\nabla_\alpha\nabla_\beta-g_{\alpha\beta}\Box)f_R\right],
\end{eqnarray}
where $T^{(m)}_{\alpha\beta}$ represents the energy momentum tensor for ordinary
matter.

The system that we have chosen for analysis is cylindrically symmetric object which consists on
a timelike three dimensional boundary surface $\Sigma$. Under consideration boundary surface $\Sigma$
constitutes two regions termed as interior and exterior region. Interior region inside the boundary \cite{32} is
\begin{equation}\label{7}
ds^2_-=A^2(t,r)dt^{2}-B^2(t,r)dr^{2}-C^2(t,r)d\phi^{2}-dz^{2},
\end{equation}
whereas the line element for exterior region \cite{33} can be defined with the help of
following diagonal form:
\begin{equation}\label{8}
ds^2_+=\left(-\frac{2M}{r}\right)d\nu^2+2drd\nu-r^2(d\phi^2+\zeta^{2}
dz^{2}),
\end{equation}
where $'\nu'$ is the retarded time, $'M'$ is total
gravitating mass and $\zeta$ represents an arbitrary constant. The fluid can be indicated through
the following configurations of the mathematical form \cite{34}
\begin{equation}\label{9}
T^{(m)}_{\alpha\beta}=(\mu+p_r)V_{\alpha}V_{\beta}-p_rg_{\alpha\beta}+(p_z-p_r)S_{\alpha}S_{\beta}+(p_\phi-p_r)K_{\alpha}K_{\beta},
\end{equation}
where $\mu$ represent energy density, $p_r$, $p_\phi$ and $p_z$ are principal stresses,
while $V_{\beta}$, $S_{\beta}$ and $K_{\beta}$ denote four-velocity and four-vectors respectively.
Under co-moving relative motion, these four-vectors and four-velocity are defined as
\begin{equation}\label{10}
V_{\beta}=A\delta^{0}_{\beta},\quad
S_{\beta}=S\delta^3_{\beta},\quad
K_\beta=C\delta^{2}_\beta,
\end{equation}
and satisfy the following relations:
\begin{eqnarray}\nonumber
V^{\beta}V_{\beta}=-1,\quad
K^{\beta}K_{\beta}=S^{\beta}S_\beta=1,\quad
V^{\beta}K_\beta=S^{\beta}K_\beta=V^{\beta}S_\beta=0.
\end{eqnarray}
The expansion scalar $\Theta$ defines the rate of change of matter distribution and is given by the
following mathematical formula
\begin{equation}\label{11}
\Theta=V^\alpha_{;\alpha
}=\frac{1}{A}\left(\frac{\dot{B}}{B}+\frac{\dot{C}}{C}\right),
\end{equation}
where `dot' indicates partial derivatives w.r.t time coordinate.

The Ricci invariant corresponding to the interior region given in
$(\ref{7})$ is
\begin{eqnarray}\label{12}
R(r,t)&=&\frac{2}{B^{2}}\left[\frac{A''}{A}+\frac{C''}{C}+\frac{A'}{A}\left(\frac{C'}{C}-\frac{B'}{B}\right)-\frac{B'C'}{BC}\right]
-\frac{2}{B^{2}}\left[\frac{\ddot{B}}{B}+\frac{\ddot{C}}{C}-\frac{\dot{A}}{A}\left(\frac{\dot{B}}{B}+\frac{\dot{C}}{C}\right)
+\frac{\dot{B}\dot{C}}{BC}\right],
\end{eqnarray}
where `prime' represents partial derivatives w.r.t radial coordinate.

The $f(R,T)$ field equations for the interior of cylindrically symmetric system are
\begin{eqnarray}\label{9*}
&&G_{00}=\frac{A^{2}}{f_R}\left[\mu+\frac{f-Rf_R}{2}+\eta_{00}\right],\quad
G_{01}=\frac{1}{f_R}\left(\dot{f'_R}-\frac{A'}{A}\dot{f_R}-\frac{\dot{B}}{B}f'_R\right),\\\label{10*}
&&G_{11}=\frac{B^{2}}{f_R}\left[P_r(1+f_T)+\mu
f_T-\frac{f-Rf_R}{2}+\eta_{11}\right],\quad
G_{22}=\frac{C^{2}}{f_R}\left[P_{\phi}(1+f_T)+\mu
f_T-\frac{f-Rf_R}{2}+\eta_{22}\right],\\\label{11*}
&&G_{33}=\frac{1}{f_R}\left[P_z(1+f_T)+\mu
f_T-\frac{f-Rf_R}{2}+\eta_{33}\right],
\end{eqnarray}
where
\begin{eqnarray}\label{12*}
\eta_{00}
&=&\frac{f_R}{B^2}-\frac{\dot{f_R}}{A^2}\left(\frac{\dot{B}}{B}+\frac{\dot{C}}{C}\right)
-\frac{f_R'}{B^2}\left(\frac{B'}{B}-\frac{C'}{C}\right),\quad
\eta_{11}
=\frac{\ddot{f_R}}{A^{2}}+\frac{\dot{f_R}}{A^{2}}\left(\frac{\dot{C}}{C}-\frac{\dot{A}}{A}\right)
-\frac{f'_R}{B^{2}}\left(\frac{A'}{A}+\frac{C'}{C}\right),\\\label{13*}
\eta_{22}&=&\frac{\ddot{f_R}}{A^{2}}-\frac{f''_R}{B^{2}}-\frac{\dot{f_R}}{A^{2}}\left(\frac{\dot{A}}{A}-\frac{\dot{B}}{B}\right)
-\frac{f'_R}{B^{2}}\left(\frac{A'}{A}-\frac{B'}{B}\right),\\\label{14*}
\eta_{33}&=&\frac{\ddot{f_R}}{A^{2}}-\frac{f''_R}{B^{2}}-\frac{\dot{f_R}}{A^{2}}\left(\frac{\dot{A}}{A}-\frac{\dot{B}}{B}-\frac{\dot{C}}{C}\right)
-\frac{f'_R}{B^{2}}\left(\frac{A'}{A}-\frac{B'}{B}+\frac{C'}{C}\right).
\end{eqnarray}
The dynamical equations are significant for the establishment of the
stability range of relativistic bodies, so we are interested in their
construction. In $f(R,T)$ framework, it is observed that the divergence of energy-momentum tensor is
non-vanishing and is found to be
\begin{eqnarray}
\nabla{^{\alpha}}T_{\alpha\beta}=\frac{f_T}{1-f_T}\left[(\Theta_{\alpha\beta}-T_{\alpha\beta})\nabla{^{\alpha}}ln
f_T-\frac{1}{2}g_{\alpha\beta}\nabla{^{\alpha}}+\nabla^{^{\alpha}}\Theta_{\alpha\beta}\right].\label{14**}
\end{eqnarray}
The divergence of ``effective energy-momentum'' tensor which is mentioned in above expression yields the
following two equations:
\begin{eqnarray}\label{B1s}
\nonumber&&\dot{\mu}\left(\frac{1-f_T+f_R f_T}{f_R(1-f_T)}\right)-\mu\frac{\dot{f_R}}{f_R}+
\frac{\dot{B}}{B}\frac{1}{f_R}(1+f_T)(\mu+P_r)+\frac{\dot{C}}{C}\frac{1}{f_R}(1+f_T)(\mu+P_\phi)
+\frac{\dot{T}}{2}\frac{f_T}{1-f_T}+H_0(r,t)=0,\\\\\nonumber
&&P'_r\left(\frac{1-f^{2}_T+2f_R
f_T}{f_R(1-f_T)}\right)+\frac{P_r}{f_R}\left(f'_T-\frac{f'_R(1+f_T)}{f_R}\right)+\mu'\frac{f_T}{f_R}+
\frac{\mu}{f_R}\left(f'_T-\frac{f_T f'_R}{f_R}\right)+\frac{A'}{A}\frac{1}{f_R}(1+f_T)(\mu+P_r)\\\label{B2s}
&&+\frac{C'}{C}\frac{1}{f_R}(1+f_T)(P_r-P_\phi)+\frac{f'_T}{1-f_T}(\mu+P_r)+\frac{f_T}{1-f_T}\frac{T'}{2}
+\frac{f_T}{1-f_T}\mu'+H_1(r,t)=0.
\end{eqnarray}
These are the required dynamical equations that will lead
to collapse equation. Here, $H_0(r,t)$ and $H_1(r,t)$ represent
extra curvature dark source terms emerging from the $f(R,T)$ gravitational
field and these quantities are given in Appendix.
The variation of physical parameters of gravitating system with the passage
of time can be observed with the help of perturbation scheme as given in the below section.

\section{$f(R,T)$ Model and Perturbation Scheme}

Perturbation scheme is a mathematical tool that is used to find approximate
solution of a differential equation. After applying perturbation scheme,
the corresponding equation can be broken into parts i.e., static and perturbed parts.
We employ this theory to analyze the effects of $f(R,T)$ model on the evolution of
celestial body under consideration. We have applied perturbation theory in such a way that initially all the
quantities are in static equilibrium but with the passage of time perturbed
quantities have both radial and time dependence.
The selection of model is very important for analysis. The $f(R, T)$ model
we have considered for evolution analysis is combination of extended Starobinsky model \cite{35} and linear term of trace $T$, written mathematically as
\begin{eqnarray}\label{m}
&&f(R, T)= R+\alpha R^2+\gamma R^n+\lambda T,
\end{eqnarray}
where $n\geq3$, $\alpha$ and $\gamma$ corresponds to the positive real
values, while $\lambda$ is coupling parameter and $\lambda T$ represent modification to
$f(R)$ gravity. Assuming $0<\varepsilon\ll1$,
the functions may be written in the following form:
\begin{eqnarray}\label{41} A(t,r)&=&A_0(r)+\varepsilon
D(t)a(r),\\\label{42} B(t,r)&=&B_0(r)+\varepsilon
D(t)b(r),\\\label{43} C(t,r)&=&C_0(r)+\varepsilon
D(t)\bar{c}(r),\\\label{44} \mu(t,r)&=&\mu_0(r)+\varepsilon
{\bar{\mu}}(t,r),\\\label{45} p_r(t,r)&=&p_{r0}(r)+\varepsilon{\bar{p_r}}(t,r),
\\\label{46}p_\phi(t,r)&=&p_{0}(r)+\varepsilon {\bar{p_\phi}}(t,r),
\\\label{47}p_z(t,r)&=&p_{0}(r)+\varepsilon {\bar{p_z}}(t,r)\\\label{48} m(t,r)&=&m_0(r)+\varepsilon {\bar{m}}(t,r),
\\\label{49'} R(t,r)&=&R_0(r)+\varepsilon D_1(t)e_1(r),\\\label{50'}
T(t,r)&=&T_0(r)+\varepsilon D_2(t)e_2(r),\\\nonumber f(R,
T)&=&[R_0(r)+\alpha R_0^2(r)+\gamma R^n(r)+\lambda T_0]+\varepsilon D_1(t)e_1(r)
\\\label{51'}&&\times[1+2\alpha R_0(r)+\gamma nR_0^{n-1}]+\varepsilon\lambda D_2(t)e_2(r),
\\\nonumber f_R&=&(1+2\alpha R_0(r)+\gamma nR_0^{n-1})+\varepsilon D_1(t)e_1(r)\\\label{52'}&&\times(2\alpha+\gamma
n(n-1)R_0^{n-2}),\\\label{53'}
f_T&=&\lambda, \\\label{54'} \Theta(t,r)&=&\varepsilon\bar{\Theta}.
\end{eqnarray}
In above equations, $R_0$ represents the static part of Ricci scalar whose value is given below
\begin{eqnarray}
R_{0}(r)=\frac{2}{B_0^{2}}\left[\frac{A''_0}{A_0}+\frac{B'_0}{B_0r}
+\frac{A'_0}{A_0}\left(\frac{1}{r}-\frac{B'_0}{B_0}\right)\right],
\end{eqnarray}
while the value of perturbed part of Ricci scalar is given as
\begin{eqnarray}
-De&=&\frac{2\ddot{D}}{A_0^{2}}\left(\frac{b}{B_0}+\frac{\bar{c}}{r}\right)+D\left[\frac{4b}{B_0^{3}}R_0-\frac{2}{B_0^{2}}\left\{\frac{a''}{A_0}-\frac{aA''_0}
{A_0^{2}}+\frac{c''}{r}-\frac{A'_0}{A_0}\left\{\left(\frac{b}{B_0}\right)'-\left(\frac{\bar{c}}{r}\right)'\right\}+\left(\frac{a}{A_0}\right)'
\left(\frac{1}{r}-\frac{B'_0}{B_0}\right)\right.\right.\\\nonumber&&-\left.\left.\frac{B'_0}{B_0}\left(\frac{\bar{c}}{r}\right)'-\frac{1}{r}
\left(\frac{b}{B_0}\right)'\right\}\right].
\end{eqnarray}
The static configuration of $f(R,T)$ field equations
(\ref{9*})-(\ref{11*}) with assumption $C_0(r)=r$ is as:
\begin{eqnarray}\label{55}
\frac{1}{B_0^{2}}\frac{B'_0}{B_0}&=&\frac{1}{Z}\left(\mu_0+\frac{\lambda T_0-\alpha R_0^{2}-\gamma
(n-1)R_0^{n}}{2}+\eta_{00}^{(s)}\right),\\\label{56}
\frac{1}{rB_0^{2}}\frac{A'_0}{A_0}&=&\frac{1}{Z}\left(P_{r0}+\lambda(P_{r0}+\mu_0)-\frac{\lambda T_0-\alpha R_0^{2}-\gamma
(n-1)R_0^{n}}{2}+\eta_{11}^{(s)}\right),\\\label{57}
\frac{1}{B_0^{2}}\frac{A'_0}{A_0}\left(\frac{A''_0}{A'_0}-\frac{B'_0}{B_0}\right)&=&\frac{1}{Z}\left(P_{\phi0}+\lambda(P_{\phi0}+\mu_0)-\frac{\lambda
T_0-\alpha R_0^{2}-\gamma (n-1)R_0^{n}}{2}+\eta_{22}^{(s)}\right),\\\label{57}
\frac{1}{B_0^{2}} \left(\frac{A''_0}{A_0}-\frac{A'_0}{rA_0}\right)-\frac{B'_0}{B_0^{3}}\left(\frac{1}{r}+\frac{A'_0}{A_0}\right)
&=&\frac{1}{Z}\left(P_{z0}+\lambda(P_{z0}+\mu_0)-\frac{\lambda T_0-\alpha R_0^{2}-\gamma
(n-1)R_0^{n}}{2}+\eta_{33}^{(s)}\right),
\end{eqnarray}
where
\begin{eqnarray}\label{58} &&
\eta_{00}^{(s)}=\frac{Z_{R_0}}{B_0^{2}}\left[R''_0-R'_0\left(\frac{B'_0}{B_0}-\frac{1}{r}\right)\right]+\frac{Z_{R_0R_0}(R'_0)^{2}}{B_0^{2}},\quad
\eta_{11}^{(s)}=-\frac{Z_{R_0}R_0'}{B_0^{2}}\left(\frac{A'_0}{A_0}+\frac{1}{r}\right),\\\nonumber
&&\eta_{22}^{(s)}=\frac{Z_{R_0}}{B_0^{2}}\left[R'_0\left(\frac{B'_0}{B_0}-\frac{A'_0}{A_0}\right)-R''_0\right]+\frac{Z_{R_0R_0}(R'_0)^{2}}{B_0^{2}},\quad
\eta_{33}^{(s)}=\frac{Z_{R_0}}{B_0^{2}}\left[R'_0\left(\frac{B'_0}{B_0}-\frac{A'_0}{A_0}-\frac{1}{r}\right)-R''_0\right]
+\frac{Z_{R_0R_0}(R'_0)^{2}}{B_0^{2}}.\\\label{59}
\end{eqnarray}
After applying the static configuration, the first dynamical
equation (\ref{B1s}) is identically satisfied, while the 2nd
equation (\ref{B2s}) has the following form:
\begin{eqnarray}\nonumber
&&\frac{1}{Z}\left[\frac{1-\lambda^{2}+2\lambda
Z}{1-\lambda}P'_{r0}-\frac{(1+\lambda)Z'}{Z}P_{r0}+\lambda\left(\mu'_0-\frac{Z'}{Z}\mu_0\right)
+(1+\lambda){\frac{A'_0}{A_0}(\mu_0+P_{r0})+\frac{1+\lambda}{r}(P_{r0}-P_{\phi0})}\right]\\\label{B1s*}&&+\frac{\lambda}{1-\lambda}\left(\frac{T'_0}{2}
+\mu'_0\right)+H_1^{(s)}=0,
\end{eqnarray}
where $H_1^{(s)}$ represents the static part of $H_1$ and is given below:
\begin{eqnarray}\label{B2s*}
H_1^{(s)}=\left[\frac{1}{Z}\left(\frac{\lambda T_0+\alpha
R_0^{2}+\gamma(n-1)R_0^{n}}{2}+\eta_{11}^{(s)}\right)\right]_{,1}+\frac{1}{Z}
\left[\frac{A'_0}{A_0}\left(\eta_{00}^{(s)}+\eta_{11}^{(s)}
\right)+\frac{1}{r}\left(\eta_{11}^{(s)}-\eta_{22}^{(s)}\right)\right].
\end{eqnarray}
After the application of perturbation scheme, the dynamical equations
(\ref{B1s}) and (\ref{B2s}) obtained through the non-zero divergence
of ``effective energy-momentum tensor'' in $f(R,T)$ gravity turn out
to be
\begin{eqnarray}\label{59*}
\nonumber&&\frac{1}{Z}\left[\frac{\dot{\bar{\mu}}(1-\lambda+\lambda
Z)}{1-\lambda}+\left\{{\mu_0
eZ_{R_0}+\frac{b}{B_0}(1+\lambda)(\mu_0+P_{\phi0})
+\frac{\bar{c}}{r}(1+\lambda)(\mu_0+P_{\phi0})}+\frac{\lambda
eZ}{2(1-\lambda)}+ZH_0^{(p_1)}\right\}\dot{D}\right]=0,\\\\\nonumber
&&\frac{\ddot{D}}{A_0^2}\left[e'-\frac{1}{Z}\left\{Z_{R_0}\left(e'-\frac{A'_0}{A_0}e-\frac{b}{B_0}R_0\right)+Z_{R_0R_0}eR'_0\right\}-
e\frac{A'_0}{A_0}\right]+\bar{P'_r}\left(\frac{1-\lambda^{2}+2\lambda
Z}{Z(1-\lambda)}\right)+\bar{\mu'}\left(\frac{\lambda-\lambda^{2}+\lambda
Z}{Z(1-\lambda}\right)+\\\nonumber&&
\frac{1+\lambda}{Z}\left(\frac{Z_{R_0}R'_0}{Z}+\frac{A'_0}{A_0}+\frac{1}{r}\right)\bar{P_r}+\bar{\mu}
\left(\frac{\lambda}{Z^{2}}Z_{R_0}R'_0+\frac{1+\lambda}{Z}\frac{A'_0}{A_0}\right)-\frac{1+\lambda}{Zr}\bar{P_\phi}
+\left[\frac{eZ_{R_0}}{Z(1-\lambda)}\left\{\left(2\lambda-\frac{1-\lambda^{2}+2\lambda Z}{Z}\right)P'_{r0}\right.\right.\\
\nonumber&&\left.\left.+\left(\lambda-\frac{\lambda-\lambda^{2}+\lambda Z}{Z}\right)\mu'_0\right\}
-\frac{2eZ_{R_0}^{2}R'_0}{Z^{3}}\left\{(1+\lambda)P_{r_0}+\lambda\mu_0\right\}+\frac{1}{Z^{2}}
(e'Z_{R_0}+eZ_{R_0R_0}R'_0)\left\{(1+\lambda)P_{r_0}+\lambda\mu_0)\right\}\right.\\
\nonumber&&\left.+\frac{1+\lambda}{Z}(\mu_0+P_{r0})
\left\{\left(\frac{a}{A_0}\right)'-\frac{eZ_{R_0}}{Z}\frac{A'_0}{A_0}\right\}
+\frac{1+\lambda}{Z}(P_{r0}-P_{\phi0})\left\{\left(\frac{\bar{c}}{r}\right)'-\frac{eZ_{R_0}}{Z}\frac{1}{r}\right\}+\frac{\lambda
e'}{2(1-\lambda)}+H_1^{(p)}\right]D=0.\\\label{59}
\end{eqnarray}
where $Z=1+2\alpha R_0+\gamma nR^{n-1}_0$, $Z_{R_0}=\frac{\partial
Z}{\partial R_0}$, $Z_{R_0R_0}=\frac{\partial^2 Z}{\partial R_0^2}$
and $H_0^{(p)}$, $H_1^{(p)}$ represent perturbed parts of $H_0$ and
$H_1$ respectively which are addressed in Appendix. It is also
assumed that $D_1=D_2=D$ and $e_1=e_2=e$.

Eliminating $\dot{\bar{\mu}}$ from perturbed equation (\ref{59*})
and integrating this with respect to ``t'', we get

\begin{eqnarray}\label{60}
&&\bar{\mu}=\frac{\lambda-1}{1-\lambda+\lambda Z}\left[\mu_0
eZ_{R_0}+\frac{b}{B_0}(1+\lambda)(\mu_0+P_{r0})+\frac{\bar{c}}{r}(1+\lambda)(\mu_0+P_{\phi0})+\frac{\lambda
eZ}{2(1-\lambda)}+ZH_0^{(p)}\right]D,\\\nonumber
\end{eqnarray}
The 2nd law of thermodynamics relates $\bar{\mu}$ and $\bar{p_r}$ as ratio of specific heat
with assumption of Harrison-Wheeler type equation of state expressed in the following expression \cite{36}
\begin{eqnarray}
\bar{p_r}&=&\Gamma\frac{p_{r0}}{\mu_0+p_{r}}\bar{\mu}.
\end{eqnarray}
The adiabatic index is a measure to recognize pressure variation
with changing density. Substituting the value of $\bar{\mu}$ given
in Eq.(\ref{59*}) in above equation, we obtain
\begin{eqnarray}\nonumber
&&\bar{P_r}=\left[\frac{\lambda^{2}-1}{1-\lambda+\lambda Z}\Gamma\left\{\frac{b}{B_0}P_{r0}+
\frac{\bar{c}}{r}P_{r0}\frac{\mu_0+P_{\phi0}}{\mu+P_{r0}}\right\}+\Gamma\frac{P_{r0}}{\mu_0+P_{r0}}\left(\frac{\lambda-1}{1-\lambda+\lambda
Z}\right)\left\{\mu_0 eZ_{R_0}+\frac{\lambda eZ}{2(1-\lambda)}+ZH_0^{(P)}\right\}\right]D,\\\label{61}
\end{eqnarray}
The expressions $\bar{P_\phi}$ and $\bar{P_z}$ can be obtained from the perturbed forms of last two field equations.
Applying perturbation on field equations and eliminating $\bar{P_\phi}$ and $\bar{P_z}$ provides
\begin{eqnarray}\label{62}
&&\bar{P_\phi}=-\ddot{D}\frac{Z}{(1+\lambda)A_0^2}\left(e+\frac{b}{r^2B_0}\right)-\frac{\lambda}{1+\lambda}\bar{\mu}+DH_2,\\\label{63}
&&\bar{P_z}=-\ddot{D}\frac{Z}{(1+\lambda)A_0^{2}}\left(e+\frac{b}{B_0}+\frac{\bar{c}}{r}\right)-\frac{\lambda}{1+\lambda}\bar{\mu}+DH_3,
\end{eqnarray}
where $H_2$ and $H_3$ are provided in Appendix.\\
After substitution of values of $\bar{\mu}, \bar{P_r}, \bar{P_\phi}$, 2nd dynamical equation takes the form
\begin{eqnarray}\nonumber
&&\frac{\ddot{D}}{A_0^2}\left[e'+e\left(\frac{1}{r}-\frac{A'_0}{A_0}\right)
-\frac{1}{Z}\left\{Z_{R_0}\left(e'-\frac{A'_0}{A_0}e-\frac{b}{B_0}R_0\right)+Z_{R_0R_0}eR'_0\right\}+
\frac{b}{B_0r^{3}}\right]+\left[\frac{\lambda}{Zr}\left[\left(\frac{\lambda-1}{1-\lambda+\lambda Z}\right)\left\{\mu_0
eZ_{R_0}\right.\right.\right.\\\nonumber&&+\left.\left.\left.\frac{b}{B_0}(1+\lambda)(\mu_0+P_{r0})+
\frac{\bar{c}}{r}(1+\lambda)(\mu_0+P_{\phi0})+\frac{\lambda
eZ}{2(1-\lambda)}+ZH_0^{(P)}\right\}\right]+\left(\frac{\lambda-\lambda^{2}+\lambda
Z}{(\lambda-1)Z}\right)\left[\left(\frac{\lambda-1}{1-\lambda+\lambda Z}\right)\right.\right.\\\nonumber&&+\left.\left.\left\{\mu_0
eZ_{R_0}\frac{b}{B_0}(1+\lambda)(\mu_0+P_{r0})
+\frac{\bar{c}}{r}(1+\lambda)(\mu_0+P_{\phi0})+\frac{\lambda
eZ}{2(1-\lambda)}+ZH_0^{(P)}\right\}\right]'+\left(\frac{\lambda}{Z^{2}}Z_{R_0}R'_0
+\frac{1+\lambda}{Z}\times\frac{A'_0}{A_0}\right)\right.\\\nonumber&&\times\left.
\left[\left(\frac{\lambda-1}{1-\lambda+\lambda Z}\right)\left\{\mu
eZ_{R_0}+\frac{b}{B_0}(1+\lambda)(\mu_0+P_{r0})+\frac{\bar{c}}{r}(1+\lambda)(\mu_0+P_{\phi0})+\frac{\lambda
eZ}{2(1-\lambda)}+ZH_0^{(P)}\right\}\right]-\Gamma\right.\\\nonumber&&\times\left.\left(\frac{1-\lambda^{2}+2\lambda
Z}{Z}\right)\left\{\frac{P_{r0}}{\mu_0+P_{r0}}\left(\frac{1}{1-\lambda+\lambda Z}\right)\left(\mu_0 eZ_{R_0}+\frac{\lambda
eZ}{2(1-\lambda)}+ZH_0^{(p)}\right)\right\}'-\Gamma\left(\frac{1-\lambda^{2}+2\lambda
Z}{Z}\right)\right.\\\nonumber&&\times\left.
\left\{\left(\frac{1+\lambda}{1-\lambda+\lambda
Z}\right)\left(\frac{b}{B_0}P_{r0}+\frac{\bar{c}}{r}\frac{\mu_0+P_{\phi0}}{\mu_0+P_{r0}}P_{r0}\right)\right\}'-\Gamma\frac{1-\lambda^{2}}{Z}
\left(\frac{Z_{R_0}R'_0}{Z}+\frac{A'_0}{A_0}+\frac{1}{r}\right)\left\{\frac{P_{r0}}{\mu_0+P_{r0}}\left(\frac{1}{1-\lambda+\lambda
Z}\right)\right.\right.\\\nonumber&&\times\left.\left.\left(\mu_0 eZ_{R_0}+\frac{\lambda
eZ}{2(1-\lambda)}+ZH_0^{(p)}\right)\right\}-\Gamma\frac{1-\lambda^{2}}{Z}
\left(\frac{Z_{R_0}R'_0}{Z}+\frac{A'_0}{A_0}+\frac{1}{r}\right)\left\{\left(\frac{1+\lambda}{1-\lambda+\lambda
Z}\right)\left(\frac{b}{B_0}P_{r0}+\frac{\bar{c}}{r}\frac{\mu_0+P_{\phi0}}{\mu_0+P_{r0}}P_{r0}\right)\right\}\right.\\\nonumber&&+\left.
\frac{eZ_{R_0}}{Z(1-\lambda)}\left\{\left(2\lambda-\frac{1-\lambda^2+2\lambda
Z}{Z}\right)P'_{r0}+\left(\lambda-\frac{\lambda-\lambda^2+2\lambda
Z}{Z}\right)\mu_0'\right\}+\frac{(1+\lambda)P_{r0}+\lambda\mu_0}{Z^{2}}\left\{e'Z_{R_0}
+eZ_{R_0R_0}R'_0\right.\right.\\\nonumber&&\left.\left.-\frac{2eZ_{R_0}^{2}R'_0}{Z}\right\}
+\frac{1+\lambda}{Z}(\mu_0+P_{r0})\left\{\left(\frac{a}{A_0}\right)'-\frac{eZ_{R_0}}{Z}\frac{A'_0}{A_0}\right\}
+\frac{1+\lambda}{Z}(P_{r0}-P_{\phi0})\left\{\left(\frac{\bar{c}}{r}\right)'
-\frac{eZ_{R_0}}{Zr}\right\}+\frac{\lambda
e'}{2(1-\lambda)}\right.\\\nonumber&&\left.-\frac{1+\lambda}{Zr}H_{2}+H_1^{(p)}\right]D=0.\\\label{63}
\end{eqnarray}
A 2nd order differential equation is obtained after some manipulation in perturbed part of
Ricci scalar, which takes the form as:
\begin{eqnarray}
\ddot{D(t)}-H_4(r)D(t)=0,
\end{eqnarray}
where $H_4$ is addressed in Appendix. Here, it is presumed that all the terms in $H_4$
are positive. The solution of above differential equation is of the form:
\begin{eqnarray}\label{63*}
D(t)=-e^{\sqrt{H_4}t}.
\end{eqnarray}
To estimate the instability range in Newtonian and post-Newtonian regimes, Eq.(\ref{63*}) can be used in Eq.(\ref{63}).
The subsections followed by this section provides the dynamical analysis in both regimes.

\subsection{Newtonian Regime}

To arrive at Newtonian-approximation, we assume $\mu_0\gg p_{r0}$, $\mu_0\gg p_{\phi0}$, $\mu_0\gg p_{z0}$,
and $A_0=1$, $B_0=1$. Insertion of these assumptions along with the equations
(55) and (56) leads to the following stability conditions
\begin{eqnarray}\nonumber
&&\Gamma<\frac{H_4U+V+\frac{\lambda-\lambda^{2}+\lambda
Z}{Z(\lambda-1)}W'+\lambda\left(\frac{1}{Zr}+\frac{Z_{R0}R'_0}{Z^{2}}\right)W
+\left\{(1+\lambda)P_{r0}+\lambda\mu_0\right\}X+\mu_0\frac{(1+\lambda)a'}{Z}+(P_{r0}-P_{\phi0})Y+H_5}{\frac{(1-\lambda^2+\lambda
Z)}{(1-\lambda)Z}J'+\frac{1+\lambda}{Z}\left(\frac{Z_{R0}R'_0}{Z}+\frac{1}{r}\right)J},\\\label{64}
\end{eqnarray}
where $H_{2N}$, $H_{0N}^{(p)}$ and $H_{1N}^{(p)}$ are the terms of $H_2$, $H_0^{(p)}$ and $H_1^{(p)}$ respectively that belong to
Newtonian-approximation, and
\begin{eqnarray}\nonumber
&&U=e'+\frac{e}{r}-\frac{1}{Z}\{Z_{R_0}(e'-bR_0)+eZ_{R_0R_0}R'_0\}+\frac{b}{r^{3}},\\\nonumber
&& V=\frac{eZ_{R_0}}{Z(1-\lambda)}\left\{\left(2\lambda
-\frac{1-\lambda^{2}+2\lambda Z}{Z}\right)P'_{r0}+\left(\lambda-\frac{\lambda-\lambda^{2}+\lambda
Z}{Z}\right)\mu'_0\right\},\\\nonumber
&&W=\frac{\mu_0(\lambda-1)}{1-\lambda+\lambda Z}\left\{eZ_{R0}+(1+\lambda)\left(b+\frac{\bar{c}}{r}\right)\right\},\quad
X=\frac{1}{Z^{2}}\left(e'Z_{R_0}+eZ_{R_0R_0}R'_0-\frac{2eZ_{R_0}^{2}R'_0}{Z}\right),\\\nonumber
&&Y=\frac{1+\lambda}{Z}\left\{\left(\frac{\bar{c}}{r}\right)'-\frac{eZ_{R_0}}{Zr}\right\},\quad
J=P_{r0}\frac{1-\lambda^{2}}{1-\lambda+\lambda Z}\left(b+\frac{\bar{c}}{r}\right).
\end{eqnarray}
The above condition describes the stability range of gravitating sources. All
terms mentioned in the above inequality are presumed in a way that whole expression on right side of
the adiabatic index $\Gamma$ remains positive. For the maintenance of the positivity, following
constraints must be satisfied
\begin{eqnarray}\label{65*}
 eZ_{R_0R_0}R'_0<Z_{R_0}(bR_0-e'),\quad \left(\frac{\bar{c}}{r}\right)'>\frac{eZ_{R_0}}{Zr},\\\label{65}
 e'Z_{R_0}+eZ_{R_0R_0}R'_0>\frac{2eZ_{R_0}^2R'_0}{Z},\quad p_{r0}>p_{\phi0}.
\end{eqnarray}
\subsection{Post-Newtonian Regime}

In this approximation, we choose
\begin{eqnarray}\nonumber
    A_0=1-\frac{m_0}{r}\quad and\quad B_0=1+\frac{m_0}{r},\quad then\\\nonumber
    \frac{A'_0}{A_0}=\frac{m_0}{r(r-m_0)},\quad \frac{B'_0}{B_0}=-\frac{m_0}{r(r+m_0)},
\end{eqnarray}
Substitution of above relations in Eq.(\ref{61}) leads to the
following inequality
\begin{eqnarray}\label{66}
\Gamma<\frac{H_4H_{8}+V+\frac{\lambda-\lambda^{2}+\lambda
Z}{(\lambda-1)Z}H'_6+\left(\frac{\lambda}{Zr}+\frac{\lambda}{Z^{2}}Z_{R_0}R'_0+\frac{m_0}{r(r-m_0)}\times\frac{1+\lambda}{Z}\right)H_6}
{\frac{1-\lambda^{2}+2\lambda
Z}{Z(1-\lambda)}\left(\frac{P_{r0}}{\mu_0+P_{r0}}H_6\right)'+\frac{1+\lambda}{Z}\left(\frac{Z_{R0}R'_0}{Z}+\frac{m_0}{r(r-m_0)}
+\frac{1}{r}\right)\left(\frac{P_{r0}}{\mu_0+P_{r0}}H_6\right)}\\\nonumber
+\frac{\left((1+\lambda)P_{r0}+\lambda\mu_0\right)X+H_7-\frac{1+\lambda}{Zr}H_{2(pN)}+H_{1(pN)}^{(p)}}{\frac{1-\lambda^{2}+2\lambda
Z}{Z(1-\lambda)}\left(\frac{P_{r0}}{\mu_0+P_{r0}}H_6\right)'+\frac{1+\lambda}{Z}\left(\frac{Z_{R0}R'_0}{Z}+\frac{m_0}{r(r-m_0)}
+\frac{1}{r}\right)\left(\frac{P_{r0}}{\mu_0+P_{r0}}H_6\right)},
\end{eqnarray}
where $H_{2pN}$, $H_{0pN}^{(p)}$ and $H_{1pN}^{(p)}$ are the terms of $H_2$, $H_0^{(p)}$ and $H_1^{(p)}$
that lie in post-Newtonian era.
\begin{eqnarray}\nonumber
H_6&=&\left(\frac{\lambda-1}{1-\lambda+\lambda Z}\right)\left(\mu_0
eZ_{R_0}+\frac{rb}{r+m_0}(1+\lambda)(\mu_0+P_{r0})+\frac{\bar{c}}{r}(1+\lambda)(\mu_0+P_{\phi0})+\frac{\lambda
eZ}{2(1-\lambda)}+ZH_{0(pN)}^{(P)}\right),\\\nonumber
H_7&=&\frac{1+\lambda}{Z}\left[(\mu_0+P_{r0})\left((\frac{ra}{r-m_0})'-\frac{eZ_{R_0}}{Z}\times\frac{m_0}{r(r-m_0)}\right)+(P_{r0}
-P_{\phi0})\left((\frac{\bar{c}}{r})'-\frac{eZ_{R_0}}{Zr}\right)\right]+\frac{\lambda e'}{2(1-\lambda)},\\\nonumber
H_8&=&\frac{r^2}{(r-m_0)^2}\left[e'+\frac{e}{r}+\frac{b}{r^2(r+m_0)}-\left\{\frac{Z_{R_0}}{Z}\left(e'-e\frac{m_0}{r(r-m_0)}
-\frac{br}{r-m_0}R'_0\right)+\frac{Z_{R_0R_0}eR'_0}{Z}\right\}
-\frac{em_0}{r(r-m_0)}\right].
\end{eqnarray}
Again, to maintain the positivity of  right side of the inequality
(\ref{66}), following conditions must be fulfilled
\begin{eqnarray}\nonumber
&&\left(\frac{ra}{r-m_0}\right)'>\frac{eZ_{R_0}}{Z}\times\frac{m_0}{r(r-m_0)},\quad \left(\frac{\bar{c}}{r}\right)'>\frac{eZ_{R_0}}{Zr},\\\nonumber
&&e'+\frac{e}{r}+\frac{b}{r^2(r+m_0)}>\frac{Z_{R_0}}{Z}\left(e'-e\frac{m_0}{r(r-m_0)}-\frac{br}{r-m_0}R'_0\right)+\frac{Z_{R_0R_0}eR'_0}{Z}
-\frac{em_0}{r(r-m_0)}.
\end{eqnarray}

\section{Summary}

The cosmological observations from recent data-sets like cosmic microwave background, clustering spectrum, weak lensing, Planck data
and supernovae type Ia unfolded that universe is expanding at an accelerated rate. Alternative gravitational theories have become a paradigm in description of gravitational interaction and its impact on expansion of universe. The alternative theories of gravity can be categorized as theories with
extra gravitational fields, extra spatial
dimension and higher derivatives.
A large number of mechanisms have been presented to interpret
the accelerated expansion of universe based on
improvement of Einstein theory.

The $f(R,T)$ theory of gravity being generalization of $f(R)$ gravity representing alternative gravitation theory constituting non-minimal curvature matter coupling has gained increasing attention in recent years. Its gravitational action includes additional scalar force (trace of energy-momentum tensor) together with the function $f(R)$ of Ricci scalar, that further modifies the gravitational interaction.
A scalar force is always appealing because it can reduce the
time of collapse, so the addition of extra scalar term of $T$ in modified Einstein-Hilbert
action provides better description of so called exotic matter.

Exploration of instability range in extended theories of gravity provide insight of gravitational interaction in current era that is expansion of universe.
In this paper, we have studied the impact of $f(R,T)$ model
on dynamical instability of cylindrically symmetric objects.
The selection of $f(R,T)$ model for dynamical analysis is
restricted to the form $f(R,T)=f(R)+\lambda T$, where $\lambda$ is
an arbitrary positive constant. The $f(R,T)$ model under consideration
constitutes combination of extended Starobinsky model
i.e., $f(R)=R+\alpha R^2+\gamma R^n$ for positive real values of $\alpha$,
$\gamma$ and trace $T$ which provides a suitable
replacement for dark source entities.
The gravitating system chosen for analysis is assumed to be filled with
anisotropic fluid in the interior.

For dynamical analysis, we have started with the construction of modified field equations within framework of $f(R,T)$ gravity for cylindrically symmetric gravitating system evolved under locally anisotropic background.
Covariant divergence of effective energy-momentum tensor is taken into account to arrive at dynamical equations.
The gravitating field equations describe a set of non-linear differential equations which are enough complex and their
solutions are still unknown. That is why, we have chosen perturbation approach to counter this problem
and considered the perturbation scheme proposed by Herrera et al.\cite{37}.
The system is assumed to be static at initial stage, then gradually enters into the non-static phase depending on radial and time coordinates constituting same time dependence parameter.

In order to count with the issue of instability in a gravitating system,
one may utilize numerical techniques or employ analytic approach.
Highly complicated non-linear onset of modified field equations can be
tackled essentially by implementing some numerical techniques and is of
great importance in numerical relativity. The numerical design of
Jeans analysis devised in \cite{57}-\cite{59} can be adopted for
the deep insight of dynamical analysis of a particular gravitating
source. However, numerical analysis may be confined to some
particular model with some specific ranges of physical parameters
and so turn out to be model dependent. We have chosen analytic
approach to discuss the dynamics of stellar evolution for a class of
models and presented general outcomes of gravitational interaction.

Linear perturbation has been applied on field equations and
dynamical equations. The expressions for energy density $\bar{\mu}$
and principal stresses $\bar{p_\phi}$ and $\bar{p_z}$ are obtained
from perturbed forms of dynamical equation (\ref{B1s}) and field
equations (\ref{10*}) and (\ref{11*}) respectively, while
$\bar{p_r}$ is extracted from the Harrison-Wheeler type equation.
Perturbed dynamical equations together with perturbed differential
equations leads to evolution equation that provides comprehensive
description of celestial body of cylindrical shape for dynamical
analysis.
Analytic description of evolving stars can be carried through
estimation of evolution equation constituting expression for
adiabatic index $\Gamma$.
The adiabatic index describes the stiffness in fluid distribution which is helpful in the estimation of instability eras
for gravitational bodies in the presence of expansion scalar.

It is significantly important to
mention here that the results of any gravitational theory must meet with the well-tested results of Newtonian (N) and
post-Newtonian (pN) theories. Although, gravitational field is
thought to be weak in N and pN regimes but testing the outcome in
these eras is of fundamental importance. Corrections to N and pN
regimes can be settled in $f(R,T)$ gravity that must be negligible or coincide with N and pN
approximations.
The adiabatic index $\Gamma$  requires positivity
of the terms for the maintenance of stability of celestial objects in both N and pN regimes.
Physical parameters involved in evolution equation are constrained to meet with the requirement of positive terms in expression for $\Gamma$ discussed in the subsections \textbf{A} and \textbf{B} of section
III.

It is
observed that the terms appearing in $\Gamma$ are less constrained
in weak field regimes as compared to the terms that appear in case of $f(R)$ gravity, thus $f(R,T)$ gravity represents a wider class of viable models. Thus Corrections to GR can be made
by assuming $\alpha\rightarrow{0}$, $\gamma\rightarrow{0}$,
$\lambda\rightarrow{0}$, while only $\lambda\rightarrow{0}$ leads to
the correction of $f(R)$ gravity.  The local isotropy can be
established by considering pressures same in $r$, $\phi$ and $z$
directions.

\renewcommand{\theequation}{A.\arabic{equation}}
\setcounter{equation}{0}
\section*{Appendix A}
\begin{eqnarray}\setcounter{equation}{1}\nonumber
H_0(r,t)&=&\left\{\frac{1}{f_R
}\left(\frac{f-Rf_R}{2}+\eta_{00}\right)\right\}_{,0}-\frac{1}{B^2}\left\{\frac{1}{f_R}\left(\dot{f'_R}-\frac{A'}{A}\dot{f_R}
-\frac{\dot{B}}{B}f'_R\right)\right\}_{,1}-{\frac{1}{B^{2}f_R}\left(\dot{f'_R}-\frac{A'}{A}\dot{f_R}
-\frac{\dot{B}}{B}f'_R\right)}\\\nonumber&&\times\left(\frac{A'}{A}-\frac{B'}{B}+\frac{C'}{C}\right)+\frac{\eta_{00}}{f_R}\left(\frac{\dot{B}}{B
}+\frac{\dot{C}}{C}\right)+\frac{\dot{B}}{B}\frac{\eta_{11}}{f_R}+\frac{\dot{C}}{C}\frac{\eta_{22}}{f_R},\\\label{B3}\\\nonumber
H_1(r,t)&=&\left\{\frac{1}{f_R
}\left(\frac{Rf_R-f}{2}+\eta_{11}\right)\right\}_{,1}-\frac{1}{A^2}\left\{\frac{1}{f_R}\left(\dot{f'_R}-\frac{A'}{A}\dot{f_R}
-\frac{\dot{B}}{B}f'_R\right)\right\}_{,0}+{\frac{1}{B^{2}f_R}\left(\dot{f'_R}-\frac{A'}{A}\dot{f_R}
-\frac{\dot{B}}{B}f'_R\right)}\\\nonumber&&\times\left(\frac{\dot{A}}{A}-\frac{\dot{B}}{B}+\frac{\dot{C}}{C}\right)
+\frac{A'}{A}\frac{1}{f_R}\left(\eta_{00}+\eta_{11}\right)+\frac{C'}{C}\frac{1}{f_R}\left(\eta_{11}-\eta_{22}\right).\\\label{B4}
\end{eqnarray}
\begin{eqnarray}\nonumber
H_0^{(p)}&=& \frac{e}{2Z^2}\left\{\lambda Z-Z_{R_0}(R_0+\alpha R_0^2+\gamma R_0^n+\lambda T_0)\right\}
-\frac{1}{B_0^{2}}\left\{\frac{1}{Z}\left\{Z_{R_0}\left(e'-\frac{A'_0}{A}e-R'_0\frac{b}{B_0}\right)+eR'_0Z_{R_0R_0}\right\}\right\}_{,1}
+\frac{1}{Z}\eta_{00}^{(p_1)}\\\nonumber&&-\frac{1}{B_0^{2}}\left\{\frac{1}{Z}\left\{Z_{R_0}\left(e'-\frac{A'_0}{A}e-R'_0\frac{b}{B_0}\right)
+eR'_0Z_{R_0R_0}\right\}\right\}\left(\frac{A'_0}{A_0}-\frac{B'_0}{B_0}+\frac{1}{r}\right)+\frac{1}{Z}\left(\frac{b}{B_0}
+\frac{\bar{c}}{r}-\frac{eZ_{R_0}}{Z}\right)\eta_{00}^{(s)}+\frac{b}{B_0}\\\nonumber&&\times
\frac{\eta^{(s)}_{11}}{Z}+\frac{\bar{c}}{r}\frac{\eta^{(s)}_{22}}{Z},\\\label{B5}\\\nonumber
H_1^{(p)}&=&\left[\frac{e}{2Z^2}\left\{Z_{R_0}(R_0+\alpha R_0^2+\gamma R_0^n+\lambda T_0)-\lambda
Z\right\}+\frac{\eta_{11}^{(p_1)}}{Z}-\frac{eZ_{R_0}}{Z^2}\eta_{11}^{(s)}\right]_{,1}+\frac{1}{Z}\left[\left\{\left(\frac{a}{A_0}\right)'
-\frac{eZ_{R_0}}{Z}\times\frac{A'_0}{A_0}\right\}\right.\\\nonumber&&\times\left.\left(\eta_{00}^{(s)}+\eta_{11}^{(s)}\right)+\frac{A'_0}{A_0}
\left(\eta_{00}^{(p_1)}+\eta_{11}^{(p_1)}\right)+\left\{\left(\frac{\bar{c}}{r}\right)'-\frac{eZ_{R_0}}{Z}\times\frac{1}{r}\right\}\left(\eta_{11}^{(s)}-
\eta_{22}^{(s)}\right)+\frac{1}{r}\left(\eta_{11}^{(p_1)}-\eta_{22}^{(p_1)}\right)\right].
\end{eqnarray}
\begin{eqnarray}\nonumber
H_{2}&=&\frac{1}{1+\lambda}\left[\frac{Z}{A_0B_0r^2}\left\{\frac{2bA'_0B'_0}{B_0^2}+\left(\frac{a'}{B_0}\right)'-\frac{(A_0b)'}{B_0^2}-\left(\frac{a}{A_0}
+\frac{b}{B_0}\right)\left(\frac{A'_0}{B_0}\right)'+\frac{1}{Z}\left(eZ_{R_0}-\frac{2\bar{c}Z}{r}\right),
\right.\right.\\\label{B3}&&\times\left.\left.\left(\frac{A'_0}{B_0}\right)'\right\}+\frac{e}{2}(\lambda-R_0Z_{R_0})-\eta^{(p_1)}_{22}\right],\\\nonumber
H_{3}&=&\frac{1}{1+\lambda}\left[\frac{Z}{A_0B_0}\left\{\frac{2bA'_0B'_0}{B_0^2}+\left(\frac{a'}{B_0}\right)'-
\frac{(A_0b)'}{B_0^2}-\frac{1}{B_0^2}\left(\frac{a}{A_0}
+\frac{b}{B_0}\right)\left(\frac{A'_0}{B_0}\right)'+\frac{eZ_{R_0}}{Z}\left(\frac{A'_0}{B_0}\right)'\right\}
-\frac{1}{rB_0^2}\right.\\\nonumber&&\times\left.\left\{eZ_{R_0}\frac{(A_0B_0)}{A_0B_0}'
+\frac{1}{B_0}\left(b'-\frac{3bB'_0}{B_0}\right)-r\left(\frac{A'_0}{A_0}-\frac{B'_0}{B_0}\right)
\left(\frac{\bar{c}}{r}\right)'-\bar{c}''-\left(\frac{a}{A_0}\right)'+\frac{2b}{B_0}\frac{A'_0}{A_0}\right\}+\frac{e}{2}
\right.\\\label{B4}&&\times\left.(\lambda-R_0Z_{R_0})
-\eta^{(p_1)}_{33}\right],\\\nonumber
H_{4}&=&\frac{A_0^2B_0r}{br+B_0\bar{c}}\left[\frac{1}{B_0^2}\left\{\frac{a''}{A_0}-\frac{A''_0a}{A_0^2}+\frac{\bar{c}''}{r}
-\frac{B'_0}{B_0}\left(\frac{\bar{c}}{r}\right)'-\frac{1}{r}\left(\frac{b}{B_0}\right)'+\left(\frac{a}{A_0}\right)'\left(\frac{1}{r}
-\frac{B'_0}{B_0}\right)+\frac{A'_0}{A_0}\left\{\left(\frac{\bar{c}}{r}\right)'
-\left(\frac{b}{B_0}\right)'\right\}\right.\right.\\\label{B5}&&-\left.\left.\frac{2b}{B_0}R_0\right\}-\frac{e}{2}\right],\\\nonumber
H_{5}&=&\left(\frac{\lambda-1}{1-\lambda+\lambda Z}\right)\left(\frac{\lambda}{Zr}+\frac{\lambda Z_{R_0}R'_0}{Z_0^2}\right)
\left(\frac{\lambda eZ}{2(1-\lambda)}+ZH_{0N}^{(p)}\right)+\frac{\lambda-\lambda^2+\lambda Z}{Z}\left\{\frac{1}{1-\lambda+\lambda
Z}
\left(\frac{\lambda eZ}{2(1-\lambda)}+ZH_{0N}^{(p)}\right)\right\}\\\nonumber&&+\frac{\lambda
e'}{2(1-\lambda)}+H_{1N}^{(p)}-\frac{1+\lambda}{Zr}H_{2N}.\\\label{B6}
\end{eqnarray}

\end{document}